\documentclass[pre,twocolumn,showpacs,superscriptaddress]{revtex4}
\usepackage{amsmath,amssymb}
\usepackage{graphicx,epsfig,color}
\usepackage{times}
\newcommand{\bra}[1] { \langle #1 |}
\newcommand{\ket}[1] {| #1 \rangle}

\newcommand{\aver}[1] {\left \langle #1 \right \rangle}
\renewcommand{\Im}{\textrm{Im}}
\renewcommand{\Re}{\textrm{Re}}
\newcommand{\Heff}{\mathcal{H}_{\mathrm{eff}}}
\newcommand{\Ppf}{\mathcal{P}^{\mathrm{pf}}}
\newcommand{\Pgoe}{\mathcal{P}^{\mathrm{goe}}}

\begin{document}

\title{Statistics of resonance states in open chaotic systems: A perturbative approach}

\author{Charles Poli}
 \affiliation{Laboratoire de Physique de la Mati\`ere Condens\'ee, CNRS UMR 6622, Universit\'e de Nice-Sophia Antipolis, 06108 Nice cedex 2, France}
\author{Dmitry V. Savin}
 \affiliation{Department of Mathematical Sciences, Brunel University, Uxbridge, UB8 3PH, United Kingdom}
\author{Olivier Legrand}
\author{Fabrice Mortessagne}
 \affiliation{Laboratoire de Physique de la Mati\`ere Condens\'ee, CNRS UMR 6622, Universit\'e de Nice-Sophia Antipolis, 06108 Nice cedex 2, France}

\published{5 October 2009 in\ \ \texttt{Phys.\;Rev.\;E \textbf{80},\;046203\;(2009)}}

\begin{abstract}
We investigate the statistical properties of the complexness parameter which characterizes uniquely complexness (nonorthogonality) of resonance eigenstates of open chaotic systems. Specifying to the regime of weakly overlapping resonances, we apply the random matrix theory to the effective Hamiltonian formalism and derive analytically the probability distribution of the complexness parameter for two statistical ensembles describing the systems invariant under time reversal. For those with rigid spectra, we consider a Hamiltonian characterized  by a picket-fence spectrum without spectral fluctuations. Then, in the more realistic case of a Hamiltonian described by the Gaussian orthogonal ensemble, we reveal and discuss the role of spectral fluctuations.
\end{abstract}

\maketitle

\section{Introduction}

In the domain of wave or quantum chaos \cite{Sto99}, open systems are currently actively investigated both from experimental and theoretical points of view (see Refs. \cite{Kuh05,Fyo05} for recent reviews). Openness may be due to various physical mechanisms such as bulk absorption, coupling to the environment through physical channels as well as dissipative or radiative boundary conditions. Whatever the mechanism, openness results in spectral broadening ranging from the perturbative regime of non-overlapping (isolated) resonances to the so-called Ericsson regime of strong overlap. These mechanisms and their related spectral effects have been experimentally studied in various context: in microwave cavities \cite{Men03,Bar05,Dietz06,Hem06,Kuhl08}, in optical microcavities \cite{Fang07,Leb07,Rex02}, and in elastodynamics \cite{Lobkis00,Lobkis03}.

The most salient feature of open systems is the set of resonances which are quasibound states embedded in the continuum. A natural way to address them analytically is via the energy-dependent scattering matrix, $S(E)$. Following the Heidelberg approach \cite{Verb85}, the poles (i.e., resonances) of the $S$ matrix turn out to be the complex eigenvalues of an effective non-Hermitian Hamiltonian $\Heff$, whereas the bi-orthogonal eigenvectors of the latter determine the corresponding resonance states (quasimodes). Universal properties of resonance scattering in the chaotic regime can then be analyzed by applying random matrix theory (RMT) that amounts to replacing the actual non-Hermitian Hamiltonian with an RMT ensemble of the appropriate symmetry class \cite{Guh98}. The main advantage of such an approach is that it treats on equal footing both the spectral and scattering characteristics of open chaotic systems as well as that it is flexible enough to incorporate other imperfections of the system, e.g., disorder and losses \cite{Fyo05}.

By now, complex eigenvalues of such non-Hermitian random matrices have been studied quite
systematically \cite{Fyo97,Fyo99,Fyo03}. However, the statistical properties of the corresponding (left and right) eigenvectors are less understood. Quite a substantial progress in this direction has been achieved by Schomerus \emph{et al.} \cite{Scho00}, who studied mainly the systems with broken time-reversal symmetry. Other analytical results for a few  physically interesting particular cases have also been reported in the literature recently \cite{Meh01,Fyo02,Bro03,Sav06}. Components of eigenvectors appear as residues of the $S$ matrix at resonance positions and the understanding of their properties is thus important for many applications. For example, nonorthogonality of resonance eigenstates yields the enhancement (the so-called Petermann factor) of the line width of a lasing mode in open resonators  \cite{Scho00} and influences branching ratios of nuclear cross-sections \cite{Sok97ii,Oko03}. It features also in the particle escape from the scattering region \cite{Sav97} as well as in dissipative quantum chaotic maps \cite{Kea08}.

This paper focuses on spectral and eigenvector statistics of such non-Hermitian random matrices describing effective  Hamiltonians of open chaotic wave systems whose closed limit displays time-reversal symmetry (TRS). In this case, the quasimodes correspond to the complex-valued eigenvectors of $\Heff$. To characterize this complexness, it is convenient to introduce \cite{Lobkis00,Pnini96} the ratio of the variances of the imaginary and real parts of the eigenvector as a single statistical parameter, hereafter called the complexness parameter \cite{Sav06}. One should note that this parameter is characteristic of the degree of non-orthogonality of the complex modes and, therefore, is closely connected to the Petermann factor mentioned above \cite{Scho00}. Other studies have considered the phase rigidity, another related parameter, introduced to characterize the degree to which a general scattering wave function is complex \cite{Bro03,Bul06}. Both parameters are straightforwardly deduced from one another when the phase rigidity is calculated for a single eigenvector.  The main advantage of considering the complexness parameter is to reveal a physical connection between spatial and spectral statistics \cite{Bar05,Sav06}.

In what follows, we study the probability distribution of the complexness parameter for a generic weakly open chaotic system and its connection with the distribution of resonance widths within the RMT approach. At the first stage, we derive an expression for the complexness parameter in the weak coupling regime and establish a general relation between its average and width fluctuations. Then accounting for the essential statistical feature of spectra in chaotic systems, namely, spectral rigidity, we investigate the case of a system whose closed limit is described by a pure picket-fence spectrum. An exact analytic prediction for the probability distribution of the complexness parameter is derived, depending on  only two parameters: the number of open scattering channels and the mean resonance width. Finally, we consider the more realistic case of systems modeled by the Gaussian orthogonal ensemble (GOE). We derive an analytic expression for the probability distribution of the complexness parameter in this case and discuss the effect of spectral fluctuations.

\section{Effective Hamiltonian formalism}

\subsection{Scattering approach}

Open wave systems are commonly described using  the so-called projection formalism \cite{Mah69,Oko03}. The exterior coupling is modeled by $M$ scattering channels connected to $N$ levels of a closed system. The coupling to the environment turns modes, with a infinite life time, into resonances, with a finite life time. Being initially introduced in nuclear physics, this formalism has been later applied successfully to wave billiards \cite{Dit00} for which antennas and absorption can be described by scattering channels \cite{Bar05pre}. In this approach, the resonance part of the $S$-matrix is given by:
\begin{equation}
  S(E) = 1 - iV^{\dag}\frac{1}{E-\Heff}V ,
\end{equation}
where $V$ is the coupling matrix of size $N\times M$, the elements $V_n^c$ of this matrix  couple the $n$-th level to the $c$-th scattering channel. The poles of $S$ are given by the eigenvalues of $\Heff$. Assuming an independence of the coupling elements from the energy and neglecting direct processes \cite{Verb85}, the effective  Hamiltonian of the open systems is represented as follows:
\begin{equation}\label{heff}
  \Heff = H - \frac{i}{2}VV^{\dag} ,
\end{equation}
where $H$ is the Hamiltonian of the closed system and the anti-Hermitian part $\frac{i}{2}VV^{\dag}$ describes coupling to the channels. In the case of the systems with preserved TRS considered below, $H$ is a real symmetric matrix of size $N\times N$ and $V$ is also real. As usual, the limit $N\rightarrow \infty$ is to be finally taken.

Since $\Heff$ is a non-Hermitian operator, the eigenvalue problems $\Heff\ket{ \psi_n}= \mathcal{E}_n\ket{\psi_n}$ and $\bra{\tilde{\psi}_n}\Heff = \mathcal{E}_n \bra{\tilde{\psi}_n}$ define two sets of a priori independent eigenvectors, called right $\{\ket{\psi_n}\}$ and  left $\{\bra{\tilde{\psi}_n}\}$ eigenvectors associated to the same set of  eigenvalues $\{\mathcal{E}_n\}$. These eigenvectors form a bi-orthogonal set which satisfies conditions of orthogonality, $\bra{\tilde{\psi}_n}\psi_m\rangle = \delta_{nm}$, and completeness, $\sum_n\ket{\psi_n}\bra{\tilde{\psi}_n}=1$.
Making use of the right eigenvectors, the diagonalization of $\Heff$ then reads:
\begin{equation}\label{eig}
  \Heff \ket{ \psi_n} = (E_n - \frac{i}{2} \Gamma_n)\ket{ \psi_n}
\end{equation}
where $E_n$ and $\Gamma_n$ are, respectively, the energy and the width of the $n$-th resonance. Due to TRS present, $\Heff$ is a complex symmetric matrix; hence, the left and right eigenvectors are related by the transpose, $\bra{\tilde{\psi}_n}=(\ket{\psi_n})^T$  \cite{Sok89}.

The coupling to continuum, as described by the imaginary part of $\Heff$, turns real eigenfunctions of the closed system into complex quasimodes of its open counterpart. In order to measure their complexness, we define the complexness parameter $q_n^2$ as follows:
\begin{equation} \label{qdef}
  q_n^2 = \frac{\sum_i(\Im\; \psi_n^i)^2}{\sum_i(\Re\; \psi_n^i)^2}
\end{equation}
where $\psi_n^i$ is the $i$-th component of the eigenvector (we note that the complexness parameter can be equivalently defined by means of the left eigenvectors). It is worth noting here that in contrast to the related Petermann factor \cite{Scho00}, which is defined for a fixed value of the given resonance width, no additional constraints are imposed on (\ref{qdef}). In chaotic systems, $q_n^2$ reveals strong mode-to-mode fluctuations, which we describe through its probability distribution function to be derived below.

\subsection{Statistical assumptions}

Within the RMT approach, the universal statistical properties of closed chaotic systems with preserved TRS are described by GOE \cite{Sto99}. In this ensemble the joint probability distribution, $P(\{E_i\})$, of the levels (the eigenvalues of $H$) is induced by a Gaussian distribution of the random real symmetric $H$ with zero mean. The exact expression for $P(\{E_i\})$ is well known to have the following form:
\begin{equation}\label{jpd}
 P(\{E_i\})\propto \prod_{n>m} |E_n-E_m|\exp \Big ({-\frac{N\pi^2}{8}\sum_n E_n^2} \Big).
\end{equation}
Here, we have chosen the variance of $H$ such that it yields the mean level spacing $\Delta=1/N$ at the spectrum center, $E=0$.

The energy levels, as defined by Eq.~(\ref{jpd}), exhibit a linear level repulsion. As a result, the energy spectrum displays spectral rigidity which restrains the spectral fluctuations around the mean. This important feature can approximately be taken into account within the so-called picket-fence model of equidistantly spaced levels \cite{Mold67}. The usefulness of this model is in its simplicity that allows one to treat various resonance phenomena analytically, see, e.g., Refs.~\cite{Mold68,Sok97,Jun99}. Here, we employ this model to single out a contribution to $q_n^2$ due to fluctuations of the resonance widths.

As concerns the coupling amplitudes, the results are known to be model independent on statistical assumptions on $V_n^c$ as long as the number of open channels is small compared to that of the levels \cite{Leh95a,Leh95b}. The coupling amplitudes may be equivalently chosen as fixed \cite{Verb85} or random \cite{Sok89}. In order to preserve orthogonal invariance of $\Heff$ under (complex) orthogonal transformations \cite{Sok89}, we consider the $V_n^c$'s as real Gaussian random variables with zero mean and
\begin{equation}\label{V}
  \aver{V_n^c V_{n'}^{c'}}= (2\kappa\Delta/\pi) \delta_{nn'} \delta^{cc'}
  \equiv \sigma^2 \delta_{nn'} \delta^{cc'} \,.
\end{equation}
Henceforth, $\aver{\cdots}$ stands for the statistical averaging over the ensembles. The coupling constant $\kappa$ determines a transmission coefficient $T=1-|\aver{S}|^2=4\kappa/(1+\kappa)^2$ of the channels (assumed to be statistically equivalent). The cases of $T\ll 1$ or $T=1$ correspond, respectively, to weak  or perfect coupling. In the weak coupling regime considered below, $\kappa\ll1$, all the resonances are almost isolated and $\aver{\Gamma} \ll \Delta$.

\section{Perturbative approach}

\subsection{\label{sec:level1}Complexness parameter in the weak coupling regime}

We now derive an expression for the complexness parameter of the eigenvectors for weakly overlapping resonances. The matrix representation of $\Heff$ in an arbitrary basis  $\{\ket{n}\}$ of the Hilbert space spanned by eigenvectors of $H$ reads:
\begin{equation}
  \Heff = \sum_{n,p=1}^N \ket{n}H_{np} \bra{p}-\frac{i}{2}\sum_{n,p=1}^N \sum_{c=1}^M \ket{n}V_n^cV_p^c\bra{p}
\end{equation}
As we focus on the weak coupling regime, the imaginary part may be viewed as a perturbation of the  Hamiltonian of the closed system. The repulsion of the energy levels exhibited by the systems under investigation allows us to consider the eigenenergies of $H$ as nondegenerate. One can therefore apply first-order perturbation theory to obtain from (\ref{eig}) the eigenvalues and the eigenvectors of $\Heff$ straightforwardly. The eigenvalues read $E_n-\frac{i}{2}\Gamma_n$, where the $E_n$'s are the eigenvalues of $H$ and the widths $\Gamma_n$ are given by:
\begin{equation}\label{gam}
  \quad \Gamma_n=\sum_{c=1}^M (V_n^c)^2\,.
\end{equation}
The perturbed eigenvectors of $\Heff$ written in the eigenbasis $\{\ket{\phi_n}\}$ of $H$ are easily found as follows:
\begin{equation}
 \ket{\psi_n}=\ket{\phi_n}-i\sum_{p\neq n}\frac{\bra{\phi_p}VV^T\ket{\phi_n}}{2(E_n-E_p)}\ket{\phi_p}\,.
\end{equation}
Splitting then the real  and imaginary parts of $\ket{\psi_n}$, the complexness parameter (\ref{qdef}) of a given eigenvector reads
\begin{equation}\label{q2def}
  q^2_n=\sum_{p\neq n}\frac{\Gamma_{np}^2}{4(E_n-E_p)^2}\,,
\end{equation}
where we have introduced $\Gamma_{np}=\sum_{c=1}^MV_n^cV_p^c$. These quantities are responsible for the coupling and interference of the resonance states due to the common decay channels \cite{Sok89}.

In what follows, we study the statistical properties of the complexness parameter (\ref{q2def}) for $H$ being described by a picket-fence or belonging to GOE. It is worth noting here that expression (\ref{q2def}) is a sum of correlated random variables which, therefore, does not obey the standard central limit theorem. Statistics of a similar kind of objects appears, e.g., in the study of the parametric level dynamics (``curvature'') \cite{vonOppen1995} and in the context of interference effects in neutron scattering from compound nucleus \cite{Ber00}.

\subsection{Rescaled parameters and their statistics}

The complexness factor (\ref{q2def}) contains two contributions of distinct types, one is due to the internal levels and the other is due to the coupling matrix elements $\Gamma_{np}$. From a statistical point of view, these two are statistically independent of one another. We note, however, that the levels $E_n$ are mutually correlated. The quantities $\Gamma_{np}$'s, unlike the original amplitudes $V_n^c$, are also not statistically independent. Although their joint distribution can be found from (\ref{V}), the resulting expression is quite complicated \cite{Yu02}, being of little practical use for actual calculations in the present context.

To overcome the difficulty of averaging over the coupling amplitudes, we follow Sokolov and Zelevinsky \cite{Sok89} and treat an arbitrary matrix element $\Gamma_{np}$ as a scalar products between $M$-dimensional vectors $\mathbf{V}_n$ and $\mathbf{V}_p$ of the coupling amplitudes $\{V_l^c\}$ associated with the levels $l=n$ and $l=p$. This suggests a natural parametrization for $\Gamma_{np}$  in terms of the angles $\theta_{np}$ between the pairs of these $N$ vectors,
\begin{equation}
  \Gamma_{np} = (\mathbf{V}_n\cdot\mathbf{V}_p) = \sqrt{\Gamma_n\Gamma_p}\cos \theta_{np}.
\end{equation}
The main advantage of this representation is that the angles $\theta_{np}$ are mutually independent and also independent of $\Gamma_n$. The probability distribution of any angle (for $M\geq2$) can be easily found to be given by the expression for a solid angle in an $M$-dimensional space \cite{Sok89}:
\begin{equation}\label{pt}
  p_M(\theta)=\frac{\Gamma(M/2)}{\sqrt{\pi}\Gamma((M-1)/2)}\sin^{M-2}\theta .
\end{equation}
Note that $\Gamma_{np}=\sqrt{\Gamma_n\Gamma_p}$ at $M=1$. As concerns the lengths of these vectors, i.e., the widths (\ref{gam}), these are well-known to be independent and $\chi^2$ distributed according to
\begin{equation}\label{pg}
  P_M(\gamma) = \frac{1}{2^{M/2}\Gamma(M/2)}\gamma^{M/2-1}e^{-\gamma/2}\,.
\end{equation}
Henceforth $\gamma_n=\Gamma_n/\sigma^2$ stands for the dimensionless widths. This distribution function has the mean value $\aver{\gamma}=M$ and the variance
\begin{equation}\label{gmg2m}
  \mathrm{var}(\gamma) = 2M = \frac{2}{M}\aver{\gamma}^2\,.
\end{equation}
Thus the widths cease to fluctuate as the number of open channels grows, with the average width being kept fixed.

It is now convenient to express all the quantities in their natural units and to consider a rescaled  complexness parameter $X_n$ defined as follows:
\begin{equation}\label{X}
  \quad X_n \equiv \frac{\Delta^2}{\sigma^4}q^2_n
  = \gamma_n\sum_{p\ne n}\frac{\Delta^2Z_p}{4(E_n-E_p)^2}\,,
\end{equation}
where  we have introduced the following quantities
\begin{equation}\label{Z}
 Z_p = \gamma_p\cos^2\theta_{np}\,.
\end{equation}
$Z_p$ may be given a geometrical interpretation as (a square of) the projection of the vector $\sigma^{-1}\mathbf{V}_p$ along the direction given by the vector $\mathbf{V}_n$. These projections are statistically independent, as is obvious from the above discussion. The probability distribution of any projection follows readily from Eqs. (\ref{pt}) and (\ref{pg}). Performing an integration first over $\gamma$ and then over $\theta$ in the definition $P(Z)=\aver{\delta(Z-\gamma\cos^2\theta)}$, one finds
\begin{equation}\label{pZ}
 P(Z)=\frac{1}{\sqrt{2\pi Z}}e^{-Z/2}\,.
\end{equation}
Thus, surprisingly, the distribution of $Z_p$ is independent of $M$, being given by the Porter-Thomas law at any $M\geq1$.

\subsection{Average of $X$ and width fluctuations}

A general expression of the average value of the complexness parameter $X$ can be readily found from Eq.~(\ref{X}) by making use of the mutual statistical independence between the widths $\{\gamma_n\}$, the projections $\{Z_n\}$ and the levels $\{E_n\}$. Noting that $\aver{\gamma}=M$ and $\aver{Z}=1=\aver{\gamma}\aver{\cos^2\theta}$, one obtains
\begin{equation}\label{Xav}
  \aver{X} = Mf \,,
\end{equation}
where the factor $f$ depends on the statistical properties of the energies of the closed system only,
\begin{equation}\label{f}
    f = \aver{\sum_{p\ne n}\frac{\Delta^2}{4(E_n-E_p)^2}}.
\end{equation}

It is important to note that, generally, the nonzero values of the complexness parameter are solely due to fluctuations of the resonance widths. Indeed, in the extreme case of all the widths being equal, the anti-Hermitian part of $\Heff$ gets proportional to the unit matrix and, as a result, the complex (biorthogonal) eigenvectors become essentially real \cite{Sav06}.
It is, therefore, instructive to take this explicitly into account and, in view of relation (\ref{gmg2m}), bring Eq. (\ref{Xav}) to the form:
\begin{equation}\label{Xvar}
  \aver{X} = \frac{f}{2}\mathrm{var}(\gamma) \,.
\end{equation}
This expression relates the average complexness parameter to the natural measure of the width fluctuations, its variance.

Strong correlations between the complexness parameter and the spectral widths are already known. The proportionality between $\sqrt{\aver{X}}$ and the average value of the fluctuating part of damping was recently found experimentally in a chaotic microwave billiard at room temperature, where this was also explained heuristically using a ray picture based on the ergodic character of the wave system  \cite{Bar05}. Then this proportionality was established in \cite{Sav06} using a two-level RMT model and considering $M\gg1$ that was relevant for this experiment. Expression (\ref{Xvar}) readily provides this feature, in view of $\sqrt{\aver{X}}=\aver{\gamma}\sqrt{f/M}$, at any $N$ and $M$. On the other side, it captures fluctuation properties of the widths properly, e.g., yielding the vanishing $\aver{X}$ in the absorptive limit of many weakly coupled channels with the average total width kept fixed, due to the vanishing variance (\ref{gmg2m}). Therefore, we believe that relation (\ref{Xvar}) is a general feature of weakly open chaotic systems with non-degenerate spectrum in the perturbative regime. Figure 1 supports this suggestion through numerical simulations of the picket-fence and GOE models (with the details being given later in the next section).

\begin{figure}[t]
\begin{center}
\begin{picture}(0,0)%
\includegraphics{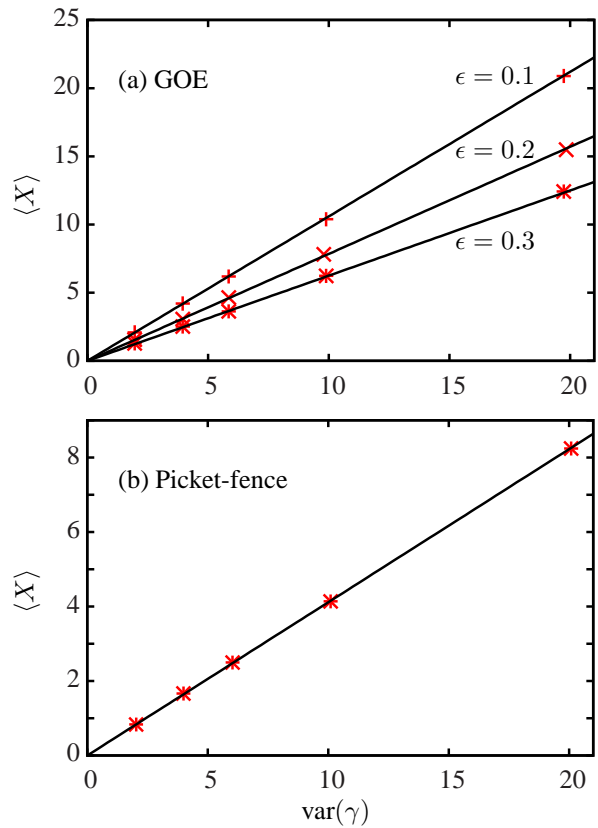}%
\end{picture}%
\setlength{\unitlength}{4144sp}%
\begingroup\makeatletter\ifx\SetFigFont\undefined%
\gdef\SetFigFont#1#2#3#4#5{%
  \reset@font\fontsize{#1}{#2pt}%
  \fontfamily{#3}\fontseries{#4}\fontshape{#5}%
  \selectfont}%
\fi\endgroup%
\begin{picture}(3513,4929)(4547,-6176)
\put(4961,-5770){\makebox(0,0)[rb]{\smash{{\SetFigFont{10}{12.0}{\familydefault}{\mddefault}{\updefault} 0}}}}
\put(4961,-5324){\makebox(0,0)[rb]{\smash{{\SetFigFont{10}{12.0}{\familydefault}{\mddefault}{\updefault} 2}}}}
\put(4961,-4878){\makebox(0,0)[rb]{\smash{{\SetFigFont{10}{12.0}{\familydefault}{\mddefault}{\updefault} 4}}}}
\put(4961,-4431){\makebox(0,0)[rb]{\smash{{\SetFigFont{10}{12.0}{\familydefault}{\mddefault}{\updefault} 6}}}}
\put(4961,-3986){\makebox(0,0)[rb]{\smash{{\SetFigFont{10}{12.0}{\familydefault}{\mddefault}{\updefault} 8}}}}
\put(4682,-2377){\rotatebox{90.0}{\makebox(0,0)[b]{\smash{{\SetFigFont{10}{12.0}{\familydefault}{\mddefault}{\updefault}{\color[rgb]{0,0,0}$\langle X\rangle$}%
}}}}}
\put(4682,-4748){\rotatebox{90.0}{\makebox(0,0)[b]{\smash{{\SetFigFont{10}{12.0}{\familydefault}{\mddefault}{\updefault}{\color[rgb]{0,0,0}$\langle X\rangle$}%
}}}}}
\put(5006,-3569){\makebox(0,0)[b]{\smash{{\SetFigFont{10}{12.0}{\familydefault}{\mddefault}{\updefault} 0}}}}
\put(5728,-3569){\makebox(0,0)[b]{\smash{{\SetFigFont{10}{12.0}{\familydefault}{\mddefault}{\updefault} 5}}}}
\put(6450,-3569){\makebox(0,0)[b]{\smash{{\SetFigFont{10}{12.0}{\familydefault}{\mddefault}{\updefault} 10}}}}
\put(7172,-3569){\makebox(0,0)[b]{\smash{{\SetFigFont{10}{12.0}{\familydefault}{\mddefault}{\updefault} 15}}}}
\put(7894,-3569){\makebox(0,0)[b]{\smash{{\SetFigFont{10}{12.0}{\familydefault}{\mddefault}{\updefault} 20}}}}
\put(5005,-5930){\makebox(0,0)[b]{\smash{{\SetFigFont{10}{12.0}{\familydefault}{\mddefault}{\updefault} 0}}}}
\put(5726,-5930){\makebox(0,0)[b]{\smash{{\SetFigFont{10}{12.0}{\familydefault}{\mddefault}{\updefault} 5}}}}
\put(6447,-5930){\makebox(0,0)[b]{\smash{{\SetFigFont{10}{12.0}{\familydefault}{\mddefault}{\updefault} 10}}}}
\put(7168,-5930){\makebox(0,0)[b]{\smash{{\SetFigFont{10}{12.0}{\familydefault}{\mddefault}{\updefault} 15}}}}
\put(7888,-5930){\makebox(0,0)[b]{\smash{{\SetFigFont{10}{12.0}{\familydefault}{\mddefault}{\updefault} 20}}}}
\put(6493,-6116){\makebox(0,0)[b]{\smash{{\SetFigFont{10}{12.0}{\familydefault}{\mddefault}{\updefault}{\color[rgb]{0,0,0}$\textrm{var}(\gamma)$}%
}}}}
\put(7682,-2701){\makebox(0,0)[rb]{\smash{{\SetFigFont{10}{12.0}{\familydefault}{\mddefault}{\updefault}{\color[rgb]{0,0,0}$\epsilon=0.3$}%
}}}}
\put(7682,-2155){\makebox(0,0)[rb]{\smash{{\SetFigFont{10}{12.0}{\familydefault}{\mddefault}{\updefault}{\color[rgb]{0,0,0}$\epsilon=0.2$}%
}}}}
\put(7682,-1711){\makebox(0,0)[rb]{\smash{{\SetFigFont{10}{12.0}{\familydefault}{\mddefault}{\updefault}{\color[rgb]{0,0,0}$\epsilon=0.1$}%
}}}}
\put(5189,-1736){\makebox(0,0)[lb]{\smash{{\SetFigFont{10}{12.0}{\familydefault}{\mddefault}{\updefault}{\color[rgb]{0,0,0}(a) GOE}%
}}}}
\put(5189,-4134){\makebox(0,0)[lb]{\smash{{\SetFigFont{10}{12.0}{\familydefault}{\mddefault}{\updefault}{\color[rgb]{0,0,0}(b) Picket-fence}%
}}}}
\put(4961,-3408){\makebox(0,0)[rb]{\smash{{\SetFigFont{10}{12.0}{\familydefault}{\mddefault}{\updefault} 0}}}}
\put(4961,-3000){\makebox(0,0)[rb]{\smash{{\SetFigFont{10}{12.0}{\familydefault}{\mddefault}{\updefault} 5}}}}
\put(4961,-2591){\makebox(0,0)[rb]{\smash{{\SetFigFont{10}{12.0}{\familydefault}{\mddefault}{\updefault} 10}}}}
\put(4961,-2184){\makebox(0,0)[rb]{\smash{{\SetFigFont{10}{12.0}{\familydefault}{\mddefault}{\updefault} 15}}}}
\put(4961,-1775){\makebox(0,0)[rb]{\smash{{\SetFigFont{10}{12.0}{\familydefault}{\mddefault}{\updefault} 20}}}}
\put(4961,-1367){\makebox(0,0)[rb]{\smash{{\SetFigFont{10}{12.0}{\familydefault}{\mddefault}{\updefault} 25}}}}
\end{picture}%
\end{center}
\caption{The average rescaled complexness parameter versus the width variance for the GOE and picket-fence models. The symbols correspond to the results of numerical simulations performed at $M=1,2,3,5,10$ (see the text for details). The linear dependence predicted by Eq.~(\ref{Xvar}) is represented by the solid line. In the GOE case (a), the proportionality factor is given by the regularized expression $f_{\epsilon}=\frac{1}{2}\int_{\epsilon}^{\infty}\!ds\,s^{-2} R_2(s)$. The results obtained with three different values of the cut-off $\epsilon$ are shown. In the picket-fence case (b), $f=\pi^2/12$ as exactly given by Eq.~(\ref{fpf}).}
\end{figure}

A remark on the proportionality factor $f$ is appropriate here. In the RMT limit $N\to\infty$, this factor may be represented as follows  $f=\frac{1}{2}\int_{0}^{\infty}\!ds\,s^{-2} R_2(s)$, where $R_2(s)$ is the two-point correlation function of the RMT. The main problem of the GOE case, already mentioned in \cite{Scho00,Sav06}, is an `infrared' logarithmical divergency of $f$ due to $R_2(s)\sim s$ at $s \rightarrow 0$. Practically, this divergence can be regularized by introducing a cut-off at small $s$, $s\geq\epsilon$, see Fig.~1. Without this cut-off the expression of the complexness parameter obtained using first order perturbation theory (\ref{X}) does not yield finite moments, thus demanding for the characterization of fluctuations of $X$ by means of its probability distribution.

\section{Distribution function}

The probability distribution function of the rescaled complexness parameter $X_n$ is defined as follows
\begin{equation}\label{PX}
  \mathcal{P}_M(X) = \aver{\delta(X-X_n)} \,,
\end{equation}
where the statistical averaging over the levels, the widths and the projections is performed with the help of Eqs.~(\ref{jpd}), (\ref{pg}), and (\ref{pZ}), respectively. In the weak coupling regime, function (\ref{PX}) depends only on the number $M$ of open channels.

It is instructive first to consider the case of the completely rigid spectrum, which may be viewed as an approximation of the GOE spectrum where the fluctuations are neglected.

\subsection{The picket-fence model}\label{sec:pf}

In this model the eigenenergies of the closed system are equally spaced, i.e. $E_n-E_{n\pm k}=\pm k\Delta$, and the eigenvector components are random Gaussian variables.  The complexness parameter is then given by
\begin{equation}\label{Xpf}
  X_n = \gamma_n \sum_{k\neq 0}\frac{Z_k}{4k^2}\,.
\end{equation}
This expression does not have any divergence problems of the GOE case, thus statistics of (\ref{Xpf}) can be also characterized by its moments. In particular, the average value is easily found to be exactly given by Eq.~(\ref{Xvar}), with the factor $f$ being
\begin{equation}\label{fpf}
 f = \sum_{k\ne 0}\frac{1}{4k^2} = \frac{\pi^2}{12}\,.
\end{equation}
Figure 1(b) illustrates the dependence $\aver{X}=\frac{\pi^2}{24}\mathrm{var}(\gamma)$.

We now derive an exact expression for the probability distribution $\Ppf_M(X)$ in the picket-fence case. First we substitute in the definition (\ref{PX}) the Fourier representation of the delta function, $\delta(X-X_n)=\int\frac{d\omega}{2\pi}e^{i\omega(X-X_n)}$, where $X_n$ is given by Eq.~(\ref{Xpf}). Then the integration over the projections $Z_k$ with the help of Eq.~(\ref{pZ}) becomes trivial, yielding
\begin{equation}\label{pdem}
 \Ppf_M(X) = \int\limits_{-\infty}^\infty \frac{d\omega}{2\pi}\;e^{i\omega X} \int \limits_{0}^\infty d\gamma \, P_M(\gamma) \prod_{k=1}^\infty \frac{1}{1+i\frac{\omega\gamma}{2k^2}} \,.
\end{equation}
The infinite product here can be evaluated explicitly \cite{Abr65}. Making use of the explicit expression (\ref{pg}) for $P_M(\gamma)$ and applying the change of variables $\gamma=2\vert z\vert^2$, Eq.~(\ref{pdem}) can then be cast in the following form:
\begin{multline}
\label{p_M(X)int}
 \Ppf_M(X) = \frac{1}{\Gamma(M/2)}\int\limits_{-\infty}^{+\infty}\frac{d\omega}{2\pi}\\
  \times \int\limits_{-\infty}^{+\infty}dz\,|z|^{M-1}\,e^{i\omega X-z^2} \frac{\sqrt{i\omega}z\pi}{\sinh(\sqrt{i\omega}z\pi)}\,.
\end{multline}

As one can easily check, this expression is properly normalized to unity. It is also worth noting that the integrand of Eq.~(\ref{p_M(X)int}) is an analytic function in $\omega$ except for the poles located on the upper part of the imaginary axis at $\omega_k=i(k/z)^2$, $k=1,2,\ldots,\infty$. This readily implies that $\Ppf_M(X)=0$ at $X<0$ identically.

\begin{figure}[t]
\begin{center}
\begin{picture}(0,0)%
\includegraphics{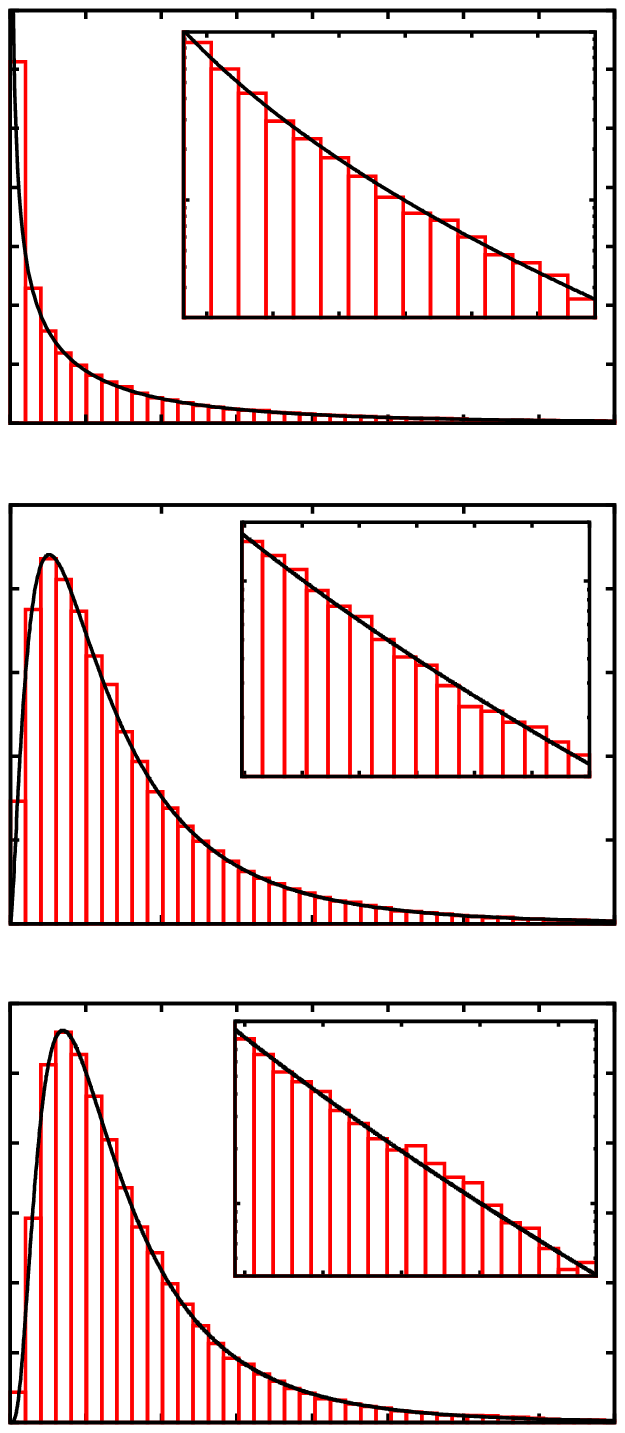}%
\end{picture}%
\setlength{\unitlength}{4144sp}%
\begingroup\makeatletter\ifx\SetFigFont\undefined%
\gdef\SetFigFont#1#2#3#4#5{%
  \reset@font\fontsize{#1}{#2pt}%
  \fontfamily{#3}\fontseries{#4}\fontshape{#5}%
  \selectfont}%
\fi\endgroup%
\begin{picture}(3484,7024)(7156,-8316)
\put(8806,-7981){\makebox(0,0)[b]{\smash{{\SetFigFont{10}{12.0}{\familydefault}{\mddefault}{\updefault} 15}}}}
\put(8115,-7981){\makebox(0,0)[b]{\smash{{\SetFigFont{10}{12.0}{\familydefault}{\mddefault}{\updefault} 5}}}}
\put(7688,-6527){\makebox(0,0)[rb]{\smash{{\SetFigFont{10}{12.0}{\familydefault}{\mddefault}{\updefault} 0.08}}}}
\put(7770,-7981){\makebox(0,0)[b]{\smash{{\SetFigFont{10}{12.0}{\familydefault}{\mddefault}{\updefault} 0}}}}
\put(7689,-5526){\makebox(0,0)[rb]{\smash{{\SetFigFont{10}{12.0}{\familydefault}{\mddefault}{\updefault} 0}}}}
\put(10533,-7981){\makebox(0,0)[b]{\smash{{\SetFigFont{10}{12.0}{\familydefault}{\mddefault}{\updefault} 40}}}}
\put(10188,-7981){\makebox(0,0)[b]{\smash{{\SetFigFont{10}{12.0}{\familydefault}{\mddefault}{\updefault} 35}}}}
\put(9843,-7981){\makebox(0,0)[b]{\smash{{\SetFigFont{10}{12.0}{\familydefault}{\mddefault}{\updefault} 30}}}}
\put(9497,-7981){\makebox(0,0)[b]{\smash{{\SetFigFont{10}{12.0}{\familydefault}{\mddefault}{\updefault} 25}}}}
\put(7689,-3993){\makebox(0,0)[rb]{\smash{{\SetFigFont{10}{12.0}{\familydefault}{\mddefault}{\updefault} 0.2}}}}
\put(9152,-7981){\makebox(0,0)[b]{\smash{{\SetFigFont{10}{12.0}{\familydefault}{\mddefault}{\updefault} 20}}}}
\put(7688,-5888){\makebox(0,0)[rb]{\smash{{\SetFigFont{10}{12.0}{\familydefault}{\mddefault}{\updefault} 0.12}}}}
\put(7688,-7806){\makebox(0,0)[rb]{\smash{{\SetFigFont{10}{12.0}{\familydefault}{\mddefault}{\updefault} 0}}}}
\put(7770,-3410){\makebox(0,0)[b]{\smash{{\SetFigFont{10}{12.0}{\familydefault}{\mddefault}{\updefault} 0}}}}
\put(8460,-3410){\makebox(0,0)[b]{\smash{{\SetFigFont{10}{12.0}{\familydefault}{\mddefault}{\updefault} 1}}}}
\put(9152,-3410){\makebox(0,0)[b]{\smash{{\SetFigFont{10}{12.0}{\familydefault}{\mddefault}{\updefault} 2}}}}
\put(9842,-3410){\makebox(0,0)[b]{\smash{{\SetFigFont{10}{12.0}{\familydefault}{\mddefault}{\updefault} 3}}}}
\put(10533,-3410){\makebox(0,0)[b]{\smash{{\SetFigFont{10}{12.0}{\familydefault}{\mddefault}{\updefault} 4}}}}
\put(7689,-1618){\makebox(0,0)[rb]{\smash{{\SetFigFont{10}{12.0}{\familydefault}{\mddefault}{\updefault} 3}}}}
\put(7689,-2157){\makebox(0,0)[rb]{\smash{{\SetFigFont{10}{12.0}{\familydefault}{\mddefault}{\updefault} 2}}}}
\put(7689,-2696){\makebox(0,0)[rb]{\smash{{\SetFigFont{10}{12.0}{\familydefault}{\mddefault}{\updefault} 1}}}}
\put(7689,-3236){\makebox(0,0)[rb]{\smash{{\SetFigFont{10}{12.0}{\familydefault}{\mddefault}{\updefault} 0}}}}
\put(7689,-4759){\makebox(0,0)[rb]{\smash{{\SetFigFont{10}{12.0}{\familydefault}{\mddefault}{\updefault} 0.1}}}}
\put(7688,-7167){\makebox(0,0)[rb]{\smash{{\SetFigFont{10}{12.0}{\familydefault}{\mddefault}{\updefault} 0.04}}}}
\put(7771,-5702){\makebox(0,0)[b]{\smash{{\SetFigFont{10}{12.0}{\familydefault}{\mddefault}{\updefault} 0}}}}
\put(8461,-5702){\makebox(0,0)[b]{\smash{{\SetFigFont{10}{12.0}{\familydefault}{\mddefault}{\updefault} 5}}}}
\put(9152,-5702){\makebox(0,0)[b]{\smash{{\SetFigFont{10}{12.0}{\familydefault}{\mddefault}{\updefault} 10}}}}
\put(9843,-5702){\makebox(0,0)[b]{\smash{{\SetFigFont{10}{12.0}{\familydefault}{\mddefault}{\updefault} 15}}}}
\put(10533,-5702){\makebox(0,0)[b]{\smash{{\SetFigFont{10}{12.0}{\familydefault}{\mddefault}{\updefault} 20}}}}
\put(8461,-7981){\makebox(0,0)[b]{\smash{{\SetFigFont{10}{12.0}{\familydefault}{\mddefault}{\updefault} 10}}}}
\put(9187,-8261){\makebox(0,0)[b]{\smash{{\SetFigFont{10}{12.0}{\familydefault}{\mddefault}{\updefault}{\color[rgb]{0,0,0}$X$}%
}}}}
\put(9327,-5009){\makebox(0,0)[b]{\smash{{\SetFigFont{10}{12.0}{\familydefault}{\mddefault}{\updefault} 14}}}}
\put(9873,-5009){\makebox(0,0)[b]{\smash{{\SetFigFont{10}{12.0}{\familydefault}{\mddefault}{\updefault} 18}}}}
\put(10417,-5009){\makebox(0,0)[b]{\smash{{\SetFigFont{10}{12.0}{\familydefault}{\mddefault}{\updefault} 22}}}}
\put(7303,-4531){\rotatebox{90.0}{\makebox(0,0)[b]{\smash{{\SetFigFont{10}{12.0}{\familydefault}{\mddefault}{\updefault}{\color[rgb]{0,0,0}$\mathcal{P}_5^\textrm{pf}(X)$}%
}}}}}
\put(7303,-2284){\rotatebox{90.0}{\makebox(0,0)[b]{\smash{{\SetFigFont{10}{12.0}{\familydefault}{\mddefault}{\updefault}{\color[rgb]{0,0,0}$\mathcal{P}_1^\textrm{pf}(X)$}%
}}}}}
\put(8727,-6021){\makebox(0,0)[rb]{\smash{{\SetFigFont{8}{9.6}{\familydefault}{\mddefault}{\updefault}$10^{-2}$}}}}
\put(8842,-7308){\makebox(0,0)[b]{\smash{{\SetFigFont{10}{12.0}{\familydefault}{\mddefault}{\updefault} 20}}}}
\put(9200,-7308){\makebox(0,0)[b]{\smash{{\SetFigFont{10}{12.0}{\familydefault}{\mddefault}{\updefault} 25}}}}
\put(9559,-7308){\makebox(0,0)[b]{\smash{{\SetFigFont{10}{12.0}{\familydefault}{\mddefault}{\updefault} 30}}}}
\put(9918,-7308){\makebox(0,0)[b]{\smash{{\SetFigFont{10}{12.0}{\familydefault}{\mddefault}{\updefault} 35}}}}
\put(10277,-7308){\makebox(0,0)[b]{\smash{{\SetFigFont{10}{12.0}{\familydefault}{\mddefault}{\updefault} 40}}}}
\put(8668,-2875){\makebox(0,0)[b]{\smash{{\SetFigFont{10}{12.0}{\familydefault}{\mddefault}{\updefault} 2}}}}
\put(8971,-2875){\makebox(0,0)[b]{\smash{{\SetFigFont{10}{12.0}{\familydefault}{\mddefault}{\updefault} 3}}}}
\put(9273,-2875){\makebox(0,0)[b]{\smash{{\SetFigFont{10}{12.0}{\familydefault}{\mddefault}{\updefault} 4}}}}
\put(9576,-2875){\makebox(0,0)[b]{\smash{{\SetFigFont{10}{12.0}{\familydefault}{\mddefault}{\updefault} 5}}}}
\put(9880,-2875){\makebox(0,0)[b]{\smash{{\SetFigFont{10}{12.0}{\familydefault}{\mddefault}{\updefault} 6}}}}
\put(10182,-2875){\makebox(0,0)[b]{\smash{{\SetFigFont{10}{12.0}{\familydefault}{\mddefault}{\updefault} 7}}}}
\put(7303,-6795){\rotatebox{90.0}{\makebox(0,0)[b]{\smash{{\SetFigFont{10}{12.0}{\familydefault}{\mddefault}{\updefault}{\color[rgb]{0,0,0}$\mathcal{P}_{10}^\textrm{pf}(X)$}%
}}}}}
\put(8841,-5009){\makebox(0,0)[b]{\smash{{\SetFigFont{10}{12.0}{\familydefault}{\mddefault}{\updefault} 10}}}}
\put(8788,-4836){\makebox(0,0)[rb]{\smash{{\SetFigFont{8}{9.6}{\familydefault}{\mddefault}{\updefault}$10^{-3}$}}}}
\put(8788,-3944){\makebox(0,0)[rb]{\smash{{\SetFigFont{8}{9.6}{\familydefault}{\mddefault}{\updefault}$10^{-2}$}}}}
\put(8509,-2246){\makebox(0,0)[rb]{\smash{{\SetFigFont{8}{9.6}{\familydefault}{\mddefault}{\updefault} $10^{-2}$}}}}
\put(8514,-1474){\makebox(0,0)[rb]{\smash{{\SetFigFont{8}{9.6}{\familydefault}{\mddefault}{\updefault} $10^{-1}$}}}}
\put(8727,-6853){\makebox(0,0)[rb]{\smash{{\SetFigFont{8}{9.6}{\familydefault}{\mddefault}{\updefault}$10^{-3}$}}}}
\end{picture}%
\end{center}
\caption{(Color online) The distribution of the rescaled complexness parameter for the picket-fence model at $M=1,5$ and 10 (top, middle and bottom, respectively). The analytical result (\ref{p_M(X)int}) is plotted in the solid line while the histograms correspond to numerics. Insets show the tail of the distribution in a semi-log scale.}
\end{figure}

The details of the subsequent calculations of $\Ppf_M(X)$ are given in Appendix \ref{AppB}. The final expression reads:
\begin{multline}
\label{p_M(X)bessel}
  \Ppf_M(X) = \frac{2\pi (\sqrt{X})^{M/2-1}}{\Gamma(M/2)}\\ \times \int\limits_{0}^{\infty}dz\, \frac{z^{M/2+1}}{\sinh(z\pi)} J_{M/2-1}(2\sqrt{X}z) \,,
\end{multline}
with $J_\nu(x)$ being the Bessel function of order $\nu$. In the case of an odd number of channels, $M=2n+1$, $n=0,1,\ldots,$ this expression can be integrated further to yield an attractively simple formula
\begin{equation}\label{p_M(X)odd}
  \Ppf_{2n+1}(X) = \frac{\sqrt{\pi}X^{n-1/2}}{2\Gamma(n+1/2)} \left(-\frac{\partial}{\partial X}\right)^{n} \frac{1}{\cosh^2(\sqrt{X})}\,.
\end{equation}
In particular, the single-channel distribution $\Ppf_1(X)$ reads
\begin{equation}\label{p_1(X)}
  \Ppf_1(X) = \frac{1}{2\sqrt{X}}\frac{1}{\cosh^2(\sqrt{X})}\,.
\end{equation}

It is interesting now to study in details the case of the large number of weakly open channels $M\gg1$. In view of the scaling (\ref{Xav}), we consider the limiting probability distribution of $x=X/M$ defined as
\begin{equation}\label{p(x)}
 p(x) = \lim_{M\to\infty} M \mathcal{P}_M(Mx) \,.
\end{equation}
Expression (\ref{p_M(X)bessel}) is actually not very convenient for evaluating this function. However, one can note that in the limit considered, the distribution $P_M(\gamma)$, Eq.~(\ref{pg}), tends to the Dirac distribution, $\delta(\gamma-M)$. Then, starting from Eq. (\ref{pdem}), the integration over $\gamma$ is trivial and the probability distribution of $x$ reads:
\begin{equation}
 p_{\mathrm{pf}}(x) = \frac{1}{2\pi}\int\limits_{-\infty}^\infty d\omega \;e^{i\omega x}\prod_{k=1}^{\infty} \frac{1}{1+i\omega/(2k^2)}
\end{equation}
Using the residue theorem, one readily gets:
\begin{equation}\label{p(x)pf}
 p_{\mathrm{pf}}(x) = 4\sum_{k=1}^{\infty}(-1)^{k+1} k^2 e^{- 2k^2x}
\end{equation}
and finally
\begin{equation}\label{jacobi}
 p_{\mathrm{pf}}(x) = -2e^{-2x}\frac{d}{dx} \vartheta_4 (0,e^{- 2x})
\end{equation}
where $\vartheta_4$ is a Jacobi theta function \cite{Abr65}.

The above analytical predictions concerning the average value of the complexness factor and its probability distribution have been checked through numerical simulations of random matrices, see Figs. 1 and 2. Numerical  simulations are based on the diagonalization of the effective  Hamiltonian (\ref{heff}) viewed as a random non-Hermitian matrix. We have considered resonances in the bulk  only, \textit{i.e.} resonances with a large number of neighbors on the left and on the right of the spectrum. This restriction is introduced to neglect the edge effects whose contribution tends to vanish as $N\rightarrow \infty$.

The picket-fence  Hamiltonian is built such that the eigenenergies are equally spaced and the eigenvectors are random Gaussian variables. This is readily done by following a procedure adapted from \cite{Gor97} where the authors used it to generate the POE ensemble. Thus, in a basis deduced from its eigenbasis through an arbitrary orthogonal transformation $O$ with random Gaussian variables, the Hamiltonian $H$ is given by:
\begin{equation}
 H = O\,\text{diag}\{ E_n\} O^t
\end{equation}
where $E_n=n/N$, such that $\Delta=1/N$, and
\begin{equation}
 \aver{O_{ij}}=0, \quad
 \aver{O_{ij}^2}=1/N
\end{equation}
Statistics were performed with 100 matrices of size 1000$\times$1000. In order to make the calculated distributions insensitive to edge effects, 100 levels at each end of the spectrum were discarded. In all the simulations the mean spectral width is kept fixed and equal to $\aver{\Gamma}/\Delta=10^{-2}$.

\subsection{The GOE model}

The probability distribution in the GOE case can be found by making use of group integration methods and results obtained in \cite{Scho00}. Outlining the details of the computation in Appendix \ref{AppC}, we state the final result here:
\begin{equation}
\label{pXgoe}
 \Pgoe_M(X)=\frac{\pi^2M}{24X^2}\frac{1+\pi^2(3+M)/(4X)}{[1+\pi^2/(4X)]^{M/2+2}}.
\end{equation}

\begin{figure}[t]
\begin{center}
\begin{picture}(0,0)%
\includegraphics{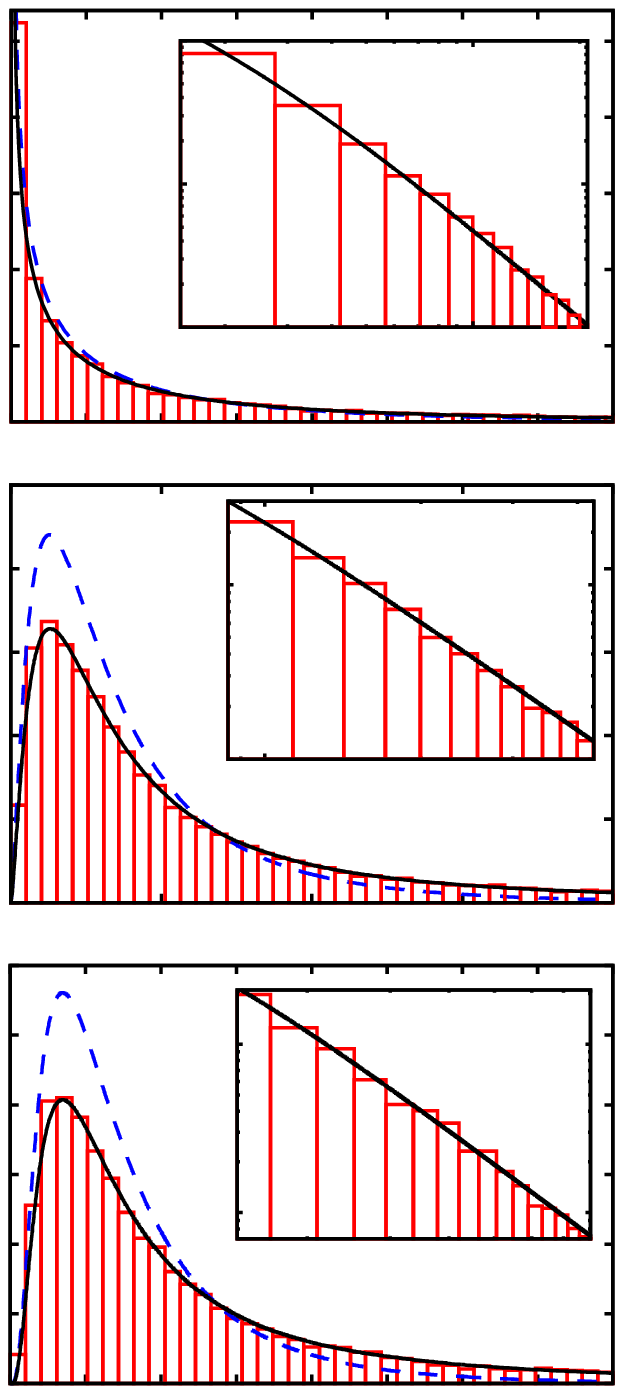}%
\end{picture}%
\setlength{\unitlength}{4144sp}%
\begingroup\makeatletter\ifx\SetFigFont\undefined%
\gdef\SetFigFont#1#2#3#4#5{%
  \reset@font\fontsize{#1}{#2pt}%
  \fontfamily{#3}\fontseries{#4}\fontshape{#5}%
  \selectfont}%
\fi\endgroup%
\begin{picture}(3476,6768)(2195,-7794)
\put(2726,-6087){\makebox(0,0)[rb]{\smash{{\SetFigFont{10}{12.0}{\familydefault}{\mddefault}{\updefault} 0.08}}}}
\put(2726,-5449){\makebox(0,0)[rb]{\smash{{\SetFigFont{10}{12.0}{\familydefault}{\mddefault}{\updefault} 0.12}}}}
\put(2731,-2963){\makebox(0,0)[rb]{\smash{{\SetFigFont{10}{12.0}{\familydefault}{\mddefault}{\updefault} 0}}}}
\put(2731,-2266){\makebox(0,0)[rb]{\smash{{\SetFigFont{10}{12.0}{\familydefault}{\mddefault}{\updefault} 1}}}}
\put(2731,-1570){\makebox(0,0)[rb]{\smash{{\SetFigFont{10}{12.0}{\familydefault}{\mddefault}{\updefault} 2}}}}
\put(4925,-2664){\makebox(0,0)[b]{\smash{{\SetFigFont{10}{12.0}{\familydefault}{\mddefault}{\updefault}10}}}}
\put(5406,-2664){\makebox(0,0)[b]{\smash{{\SetFigFont{10}{12.0}{\familydefault}{\mddefault}{\updefault}20}}}}
\put(3788,-2664){\makebox(0,0)[b]{\smash{{\SetFigFont{10}{12.0}{\familydefault}{\mddefault}{\updefault}2}}}}
\put(4162,-6826){\makebox(0,0)[b]{\smash{{\SetFigFont{10}{12.0}{\familydefault}{\mddefault}{\updefault}{\color[rgb]{0,0,0}20}%
}}}}
\put(4817,-6826){\makebox(0,0)[b]{\smash{{\SetFigFont{10}{12.0}{\familydefault}{\mddefault}{\updefault}{\color[rgb]{0,0,0}40}%
}}}}
\put(5192,-6826){\makebox(0,0)[b]{\smash{{\SetFigFont{10}{12.0}{\familydefault}{\mddefault}{\updefault}{\color[rgb]{0,0,0}60}%
}}}}
\put(5467,-6826){\makebox(0,0)[b]{\smash{{\SetFigFont{10}{12.0}{\familydefault}{\mddefault}{\updefault}{\color[rgb]{0,0,0}80}%
}}}}
\put(3972,-4656){\makebox(0,0)[b]{\smash{{\SetFigFont{10}{12.0}{\familydefault}{\mddefault}{\updefault} 10}}}}
\put(4685,-4656){\makebox(0,0)[b]{\smash{{\SetFigFont{10}{12.0}{\familydefault}{\mddefault}{\updefault} 20}}}}
\put(5101,-4656){\makebox(0,0)[b]{\smash{{\SetFigFont{10}{12.0}{\familydefault}{\mddefault}{\updefault} 30}}}}
\put(5413,-4656){\makebox(0,0)[b]{\smash{{\SetFigFont{10}{12.0}{\familydefault}{\mddefault}{\updefault} 40}}}}
\put(3765,-4492){\makebox(0,0)[rb]{\smash{{\SetFigFont{8}{9.6}{\familydefault}{\mddefault}{\updefault}$10^{-3}$}}}}
\put(3765,-3694){\makebox(0,0)[rb]{\smash{{\SetFigFont{8}{9.6}{\familydefault}{\mddefault}{\updefault}$10^{-2}$}}}}
\put(3153,-7515){\makebox(0,0)[b]{\smash{{\SetFigFont{10}{12.0}{\familydefault}{\mddefault}{\updefault} 5}}}}
\put(3498,-7515){\makebox(0,0)[b]{\smash{{\SetFigFont{10}{12.0}{\familydefault}{\mddefault}{\updefault} 10}}}}
\put(3843,-7515){\makebox(0,0)[b]{\smash{{\SetFigFont{10}{12.0}{\familydefault}{\mddefault}{\updefault} 15}}}}
\put(4187,-7515){\makebox(0,0)[b]{\smash{{\SetFigFont{10}{12.0}{\familydefault}{\mddefault}{\updefault} 20}}}}
\put(4532,-7515){\makebox(0,0)[b]{\smash{{\SetFigFont{10}{12.0}{\familydefault}{\mddefault}{\updefault} 25}}}}
\put(4876,-7515){\makebox(0,0)[b]{\smash{{\SetFigFont{10}{12.0}{\familydefault}{\mddefault}{\updefault} 30}}}}
\put(5220,-7515){\makebox(0,0)[b]{\smash{{\SetFigFont{10}{12.0}{\familydefault}{\mddefault}{\updefault} 35}}}}
\put(5564,-7515){\makebox(0,0)[b]{\smash{{\SetFigFont{10}{12.0}{\familydefault}{\mddefault}{\updefault} 40}}}}
\put(2809,-7515){\makebox(0,0)[b]{\smash{{\SetFigFont{10}{12.0}{\familydefault}{\mddefault}{\updefault} 0}}}}
\put(4192,-7739){\makebox(0,0)[b]{\smash{{\SetFigFont{10}{12.0}{\familydefault}{\mddefault}{\updefault}{\color[rgb]{0,0,0}$X$}%
}}}}
\put(3500,-5300){\makebox(0,0)[b]{\smash{{\SetFigFont{10}{12.0}{\familydefault}{\mddefault}{\updefault} 5}}}}
\put(4187,-5300){\makebox(0,0)[b]{\smash{{\SetFigFont{10}{12.0}{\familydefault}{\mddefault}{\updefault} 10}}}}
\put(4878,-5300){\makebox(0,0)[b]{\smash{{\SetFigFont{10}{12.0}{\familydefault}{\mddefault}{\updefault} 15}}}}
\put(5564,-5300){\makebox(0,0)[b]{\smash{{\SetFigFont{10}{12.0}{\familydefault}{\mddefault}{\updefault} 20}}}}
\put(2729,-5163){\makebox(0,0)[rb]{\smash{{\SetFigFont{10}{12.0}{\familydefault}{\mddefault}{\updefault} 0}}}}
\put(2729,-4399){\makebox(0,0)[rb]{\smash{{\SetFigFont{10}{12.0}{\familydefault}{\mddefault}{\updefault} 0.1}}}}
\put(2729,-3635){\makebox(0,0)[rb]{\smash{{\SetFigFont{10}{12.0}{\familydefault}{\mddefault}{\updefault} 0.2}}}}
\put(2726,-7361){\makebox(0,0)[rb]{\smash{{\SetFigFont{10}{12.0}{\familydefault}{\mddefault}{\updefault} 0}}}}
\put(2726,-6722){\makebox(0,0)[rb]{\smash{{\SetFigFont{10}{12.0}{\familydefault}{\mddefault}{\updefault} 0.04}}}}
\put(2383,-6348){\rotatebox{90.0}{\makebox(0,0)[b]{\smash{{\SetFigFont{10}{12.0}{\familydefault}{\mddefault}{\updefault}{\color[rgb]{0,0,0}$\mathcal{P}_{10}^\textrm{goe}(X)$}%
}}}}}
\put(2813,-3092){\makebox(0,0)[b]{\smash{{\SetFigFont{10}{12.0}{\familydefault}{\mddefault}{\updefault} 0}}}}
\put(3500,-3092){\makebox(0,0)[b]{\smash{{\SetFigFont{10}{12.0}{\familydefault}{\mddefault}{\updefault} 1}}}}
\put(4189,-3092){\makebox(0,0)[b]{\smash{{\SetFigFont{10}{12.0}{\familydefault}{\mddefault}{\updefault} 2}}}}
\put(4877,-3092){\makebox(0,0)[b]{\smash{{\SetFigFont{10}{12.0}{\familydefault}{\mddefault}{\updefault} 3}}}}
\put(5565,-3092){\makebox(0,0)[b]{\smash{{\SetFigFont{10}{12.0}{\familydefault}{\mddefault}{\updefault} 4}}}}
\put(2362,-4190){\rotatebox{90.0}{\makebox(0,0)[b]{\smash{{\SetFigFont{10}{12.0}{\familydefault}{\mddefault}{\updefault}{\color[rgb]{0,0,0}$\mathcal{P}_5^\textrm{goe}(X)$}%
}}}}}
\put(2342,-2039){\rotatebox{90.0}{\makebox(0,0)[b]{\smash{{\SetFigFont{10}{12.0}{\familydefault}{\mddefault}{\updefault}{\color[rgb]{0,0,0}$\mathcal{P}_1^\textrm{goe}(X)$}%
}}}}}
\put(3559,-2520){\makebox(0,0)[rb]{\smash{{\SetFigFont{8}{9.6}{\familydefault}{\mddefault}{\updefault}$10^{-3}$}}}}
\put(3559,-1865){\makebox(0,0)[rb]{\smash{{\SetFigFont{8}{9.6}{\familydefault}{\mddefault}{\updefault}$10^{-2}$}}}}
\put(3559,-1210){\makebox(0,0)[rb]{\smash{{\SetFigFont{8}{9.6}{\familydefault}{\mddefault}{\updefault}$10^{-1}$}}}}
\put(3822,-6565){\makebox(0,0)[rb]{\smash{{\SetFigFont{10}{12.0}{\familydefault}{\mddefault}{\updefault}$10^{-3}$}}}}
\put(3822,-5795){\makebox(0,0)[rb]{\smash{{\SetFigFont{10}{12.0}{\familydefault}{\mddefault}{\updefault}$10^{-2}$}}}}
\put(2811,-5300){\makebox(0,0)[b]{\smash{{\SetFigFont{10}{12.0}{\familydefault}{\mddefault}{\updefault} 0}}}}
\end{picture}%
\end{center}
\caption{(Color online) The distribution of the rescaled complexness parameter for the GOE model at $M=1,5$ and 10 (top, middle and bottom, respectively). The analytical result (\ref{pXgoe}) is shown in the solid line and compared to that (\ref{p_M(X)int}) of the picket-fence case (dashed line) while the histograms correspond to numerical simulations. Insets show the tail of the distribution in a log-log scale.}
\end{figure}

To check our findings, the same kind of numerical simulations as in the picket-fence model have been performed. The closed  Hamiltonian $H$ now belongs to GOE, its elements being defined by their first two moments:
\begin{equation}
\aver{H_{ij}}=0, \quad
\aver{H_{ij}^2}=
    \begin{cases}
        4/(N \pi^2), & \; \; i=j \\
        2/(N \pi^2), & \; \; i\neq j,
    \end{cases}
\end{equation}
where $N$ is the size of the matrix. Like in the picket-fence case, the normalization is chosen such that $\Delta=1/N$.  Statistics were obtained with 150 matrices of size 1000$\times$1000. Only levels near $E=0$ for which spacings deviate less than 5\% from $\Delta$ were kept. The agreement between numerical and analytical results is flawless, as shown in Fig. 3.

The comparison between the probability distribution of $X$ in the picket-fence model and for GOE illustrates the effects of the fluctuations of the spectrum on the complexness parameter.
The maximum of both distributions are close to each other. This is mainly due to the spectrum rigidity in both ensembles. But at large $X$ the statistical weight is larger for GOE than for the picket-fence model. This difference is introduced by the behavior of the levels at small distance: the spacing of two eigenenergies can be very small, the corresponding contribution to the complexness parameter is large, then the tail of $\mathcal{P}_M(X)$ is larger for GOE than for the picket-fence model. This feature is most explicitly seen by comparing the corresponding limiting distributions at $M\gg1$. The distribution (\ref{p(x)}) of $x$ is easily obtained from Eq. (\ref{pXgoe}) and reads
\begin{equation}\label{p(x)goe}
 p_{\mathrm{goe}}(x) = \frac{\pi^2}{24x^2} \Big(1+\frac{\pi^2}{4x}\Big) \exp \left(-\frac{\pi^2}{8x}\right)
\end{equation}
In contrast to the asymptotic exponential behavior in the picket-fence case,  $p_{\mathrm{pf}}(x)\propto e^{-2x}$, see Eq.~(\ref{p(x)pf}),  the tail of the distribution (\ref{p(x)goe}) follows a power-law decay: $p_{\mathrm{goe}}(x) \propto x^{-2}$.

\section{Conclusion}

In this paper, we have studied the statistics of complex wavefunctions associated to the resonances of weakly opened wave chaotic systems with the preserved time-reversal symmetry. More specifically, in the perturbative regime, we have considered the case of the completely rigid spectra defined through the picket-fence model and that of the GOE displaying spectral fluctuations. One of the key features of this study relies on the proportionality between the average of the complexness parameter and the variance of the resonance widths, which we believe is valid for generic nondegenerate spectra. We have also derived the exact probability distribution of the complexness parameter in these two cases.

To check the validity of the present results, recent experiments in elastodynamics are available. In particular, in the case of vibrating plates, a complete knowledge of the eigenfunctions can be obtained through noninvasive measurements \cite{Xer09} even for moderate overlap of resonances. Indeed, the understanding of the statistics of eigenfunctions beyond the perturbative regime still remains an open problem. (We note that some relevant interesting  numerical results for microwave billiards with large openings were recently reported \cite{Bul07}.) Finally, one should also note that the complexness parameter may be considered as a sensitive probe of the crossover from localized to extended states in open disordered systems \cite{Van09}.

\begin{acknowledgments}
We are grateful to H.-J. Sommers for his instructive advice and help with evaluating Eq.~(\ref{p_M(X)bessel}). One of us (D.V.S.) acknowledges gratefully the generous hospitality of LPMC in Nice and the financial support of University of Nice during his stay there. The partial financial support by BRIEF grant (D.V.S.) is also acknowledged.
\end{acknowledgments}

\appendix

\section{Derivation of Eqs.~(\ref{p_M(X)bessel}) and (\ref{p_M(X)odd})} \label{AppB}

We first note that the integrand of Eq. (\ref{p_M(X)int}) is a symmetric function in $z$ that allows us to restrict the $z$-integration to the positive axis. Then we deform the contour of integration over $\omega$ from the real to imaginary axis by putting $\Omega=i\omega$. Performing after that the scaling transformations of the integration variables, first $z\to z/\sqrt{\Omega}$ and then $\Omega\to\Omega/X$, and interchanging the order of integrations over $z$ and $\Omega$, we may cast Eq.~(\ref{p_M(X)int}) in the following form
\begin{multline}
\label{app1}
  \Ppf_M(X) = \frac{2\pi X^{M/2-1}}{\Gamma(M/2)} \int\limits_{0}^{\infty}\frac{dz\, z^M}{\sinh(\pi z)} \\ {\times} \int\limits_{-i\infty}^{+i\infty}\frac{d\Omega}{2\pi i}\Omega^{-M/2}e^{\Omega-Xz^2/\Omega} \,.
\end{multline}
To calculate here the last integral over $\Omega$, we expand $e^{-Xz^2/\Omega}$ into a series and evaluate the result termwise
\begin{multline}\label{app2}
  \sum_{k=0}^{\infty}\frac{(-Xz^2)^k}{k!}\int_{-i\infty}^{+i\infty}\limits\frac{d\Omega}{2\pi i}\Omega^{-(M/2+k)}e^{\Omega} \\ = \sum_{k=0}^{\infty}\frac{[-(\sqrt{X}z)^2]^k}{k!\Gamma(\frac{M}{2}+k)}\,,
\end{multline}
where we have used $\int_{-i\infty}^{+i\infty}\frac{d\Omega}{2\pi i}\Omega^{-\nu}e^{\Omega}=1/\Gamma(\nu)$. Making now use of the well-known series representation for the Bessel function \cite{Abr65}, one can immediately recognize the r.h.s. of (\ref{app2}) to be equal to $(\sqrt{X}z)^{1-M/2}J_{M/2-1}(2\sqrt{X}z)$. Collecting all the factors together, we finally arrive at Eq.~(\ref{p_M(X)bessel}).

Further progress is possible in the case of odd $M$. It is instructive first to start with the case of $M=1$, which turns out to play the central r\^ole in this calculation. We may use the known relation $J_{-1/2}(z)=\sqrt{2z/\pi}\cos(z)/z$ in this case \cite{Abr65}, thus $(\sqrt{X}z)^{1/2}J_{-1/2}(2\sqrt{X}z)=\frac{1}{\sqrt{\pi}}\cos(2\sqrt{X}z)$, that allows us to perform the integration in Eq. (\ref{app1}) analytically:
\begin{multline}\label{app3}
   \int_{0}^{\infty}\limits dz \frac{z}{\sinh(z\pi)} \int_{-i\infty}^{+i\infty}\limits \frac{d\Omega}{2\pi i} \Omega^{-1/2}e^{\Omega-Xz^2/\Omega} \\= \int_{0}^{\infty}\limits dz \frac{z\cos(2\sqrt{X}z)}{\sinh(z\pi)\sqrt{\pi}} = \frac{1}{4\sqrt{\pi}}\frac{1}{\cosh^2(\sqrt{X})} \,.
\end{multline}
Taking now into account the (omitted) factor $2\sqrt{\pi/X}$, we obtain $\Ppf_1(X)$, Eq.~(\ref{p_1(X)}).

The general case of odd $M=2n+1$ may be reduced to that of $M=1$ considered above, if one notices that the term $z^M/\Omega^{M/2}e^{-Xz^2/\Omega}$ in the integrand of Eq.~(\ref{app1}) can be generated by a differentiation with respect to $X$ as follows:
\begin{equation}\label{app4}
   \left(\frac{z^2}{\Omega}\right)^n \frac{z}{\Omega^{1/2}} e^{-Xz^2/\Omega} = \left(-\frac{\partial}{\partial X}\right)^n \frac{z}{\Omega^{1/2}} e^{-Xz^2/\Omega}\,.
\end{equation}
Substituting this representation into Eq.~(\ref{app1}) and changing the order of the integrations and differentiation there, we see that the resulting integral is already given by Eq.~(\ref{app3}) that readily yields the expression (\ref{p_M(X)odd}) of Sec.~\ref{sec:pf}.

\section{Derivation of Eq.~(\ref{pXgoe})} \label{AppC}

We use the recent result by Schomerus \textit{et al.} \cite{Scho00}, who calculated the joint probability distribution $P(A,B)$ of
\begin{equation}
 A=\sum_{p\ne n} \frac{\alpha^2_p}{E_p-E_n}, \qquad B=\Delta \sum_{p\ne n} \frac{\alpha^2_p}{(E_p-E_n)^2},
\end{equation}
where $\{\alpha_p \}$ are the statistically independent real Gaussian variables distributed according to
\begin{equation}\label{alpha}
 p(\alpha_p^2)=\sqrt{\frac{\pi}{2\kappa\Delta \alpha_p^2}}e^{-\pi^2\alpha_p^2/(2\kappa\Delta)}
\end{equation}
and $\{E_n\}$ are taken from the GOE. They found the following expression for $P(A,B)$:
\begin{equation}\label{pAB}
 P(A,B) = \frac{\sqrt{2\pi}}{12}\frac{1+\pi^2A^2/\kappa^2}{B^{7/2}} e^{-\frac{\kappa}{2B} (1+\pi^2A^2/\kappa^2)}
\end{equation}

We note that the above expression (\ref{pAB}) was obtained in \cite{Scho00} for the particular case of one open channel. The key fact which allows us to apply this result to our $M$-channel case is the representation (\ref{X}) in terms of projections with the distribution (\ref{pZ}). The later corresponds to the Gaussian distribution (\ref{alpha}) with $\kappa=1$ and $\frac{\pi^2}{\kappa\Delta}\alpha_p^2=Z_p$, thus giving a connection $X = \frac{\pi^2}{4}\gamma B$. Correspondingly, the distribution function of $X$ in the GOE case can be found from \begin{equation}
 \Pgoe_M(X)=\aver{\delta(X-\frac{\pi^2}{4}\gamma B)}
\end{equation}
by averaging over $A,B$ and $\gamma$. Substituting the explicit form (\ref{pAB}), it is convenient first to integrate out $B$ that yields
\begin{multline}
 \Pgoe_M(X) = \frac{\sqrt{2\pi}}{12} \int\limits_0^\infty d\gamma \; P_M(\gamma) (a\gamma)^{7/2} e^{-a\gamma/2} \\
 \times \int\limits_{-\infty}^\infty dA (1+\pi^2A^2)e^{-(a\gamma/2)\pi^2A^2}\,,
\end{multline}
with $a=\pi^2/4X$. The Gaussian integration over $A$ is now straightforward and gives
$$
 \Pgoe_M(X) = \frac{\pi^2}{24\Gamma(M/2)}\frac{1}{X^2}\int\limits_0^\infty d\gamma \bigl(\frac{\gamma}{2}\bigr)^{M/2} (1+ a\gamma) e^{-(1+a)\frac{\gamma}{2} }\,,
$$
where we have substituted expression (\ref{pg}) for $P_M(\gamma)$. The remaining integration yields Eq.~(\ref{pXgoe}).


\end{document}